\documentclass[journal]{IEEEtran}
\usepackage{amsmath}
\usepackage{amsfonts}
\usepackage{amssymb}
\usepackage{multirow}
\usepackage{graphicx}
\usepackage{picins}
\usepackage[ruled, vlined]{algorithm2e}
\usepackage{url}
\usepackage{color}

\newtheorem{assump}{\textbf{Assumption}}
\newtheorem{theorem}{\textbf{Theorem}}
\newenvironment{customthm}[1]
  {\renewcommand\thetheorem{#1}\theorem}
  {\endtheorem}

\begin{document}

\title{Accelerating Federated Learning over Reliability-Agnostic Clients in Mobile Edge Computing Systems}
\author{Wentai~Wu, \IEEEmembership{Student Member,~IEEE}, Ligang~He, \IEEEmembership{Member,~IEEE}, Weiwei~Lin, and Rui Mao
\thanks{W. Wu, L. He (corresponding author, ligang.he@warwick.ac.uk) are with the Department of Computer Science, the University of Warwick. W. Lin is with the School of Computer Science and Engineering at the South China University of Technology. Rui Mao is with College of Computer Science and Software Engineering, Shenzhen University, China}}% <-this % stops a space

\maketitle

\begin{abstract}
Mobile Edge Computing (MEC), which incorporates the Cloud, edge nodes and end devices, has shown great potential in bringing data processing closer to the data sources. Meanwhile, Federated learning (FL) has emerged as a promising privacy-preserving approach to facilitating AI applications. However, it remains a big challenge to optimize the efficiency and effectiveness of FL when it is integrated with the MEC architecture. Moreover, the unreliable nature (e.g., stragglers and intermittent drop-out) of end devices significantly slows down the FL process and affects the global model's quality in such circumstances. In this paper, a multi-layer federated learning protocol called HybridFL is designed for the MEC architecture. HybridFL adopts two levels (the edge level and the cloud level) of model aggregation enacting different aggregation strategies. Moreover, in order to mitigate stragglers and end device drop-out, we introduce regional slack factors into the stage of client selection performed at the edge nodes using a probabilistic approach without identifying or probing the state of end devices (whose reliability is agnostic). We demonstrate the effectiveness of our method in modulating the proportion of clients selected and present the convergence analysis for our protocol. We have conducted extensive experiments with machine learning tasks in different scales of MEC system. The results show that HybridFL improves the FL training process significantly in terms of shortening the federated round length, speeding up the global model's convergence (by up to 12$\times$) and reducing end device energy consumption (by up to 58\%).
\end{abstract}

% keywords
\begin{IEEEkeywords}
federated learning, mobile edge computing, distributed computing, machine learning
\end{IEEEkeywords}

\section{Introduction}
\label{secI}
The rapid advance and remarkable achievements made in the development of Artificial Intelligence (AI) have drawn an unprecedented level of attention and revealed the potential of machine learning techniques. Meanwhile, the prevalence of Internet of Things (IoT) and Edge Intelligence \cite{EI1} stimulates the efforts of pushing the computation to the edge of the network (closer to where the source of data resides) for faster response and better service quality \cite{QoE2018}. With these two streams of research endeavour, it has been a major trend to empower the end devices in IoT with AI applications -- Gartner has predicted that over 80\% of enterprise IoT projects will incorporate AI components by 2022 \cite{Gartner}. Mobile Edge Computing (MEC) \cite{MEC1}\cite{MEC2}, which consists of Cloud, edge nodes and end devices, is an emerging technology that can serve as the fundamental architecture of IoT, and provides a promising architecture for sinking AI to the edge nodes \cite{EI2}. However, there are still many obstacles when it comes to the practical AI scenarios where the participants of the model training process are end devices such as cell phones, smart sensors and wearable electronics. First, although much work has shown good performance when training AI models in a cloud-centric manner using high-spec servers that hold the entire data set, it may not be feasible in many application scenarios (e.g., clinical diagnosis \cite{health2018}) nowadays due to the data privacy concerns or the administration policies that forbid moving data out of local devices. Besides, though traditional distributed machine learning techniques (e.g., \cite{dist_SGD1}\cite{dist_SGD2}) can deal with decentralized data, they require very frequent exchange of gradients and model parameters, which results in heavy network traffic and prohibitive cost of communication in the cases where the devices are connected to the cloud via wireless channels. 

Federated Learning \cite{FL}, originally proposed by Google, is a distributed machine learning protocol designed for addressing the above-mentioned problems of data privacy and communication efficiency when training from decentralized data. A typical FL process consists of multiple rounds of training, in each of which clients (i.e., end devices) perform model training on local data and the cloud aggregates local models to produce a global model using a weight-averaging algorithm called \textit{FedAvg}. As stated by McMahan et al. \cite{FL}, the key properties of FL are: i) \textbf{unbalanced data distribution}: end devices may possess variable amounts of Non-IID (Non Independent and Identically Distributed) data; ii) \textbf{massively distributed devices}: the participants can be a huge fleet of heterogeneous end devices; iii) \textbf{limited access and communication}: data access is limited to local devices while the communication between the cloud and end devices can be slow and expensive.

% hybridFL schematic overview
\begin{figure*}[htb]
    \centering
    \includegraphics[scale=0.4]{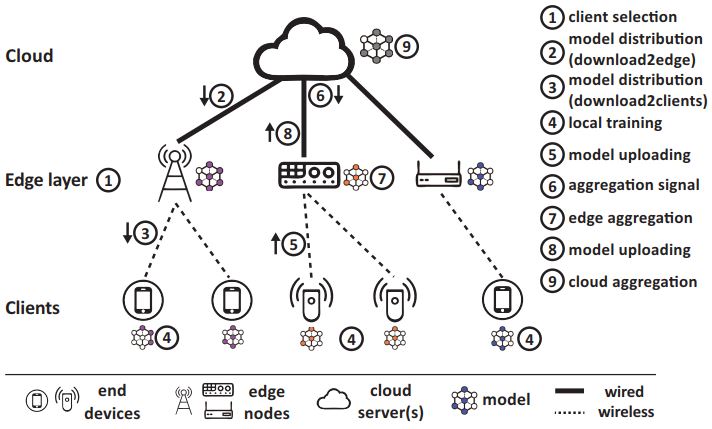}
    \caption{A schematic overview of the HybridFL protocol designed for the MEC architecture, which consists of the Cloud, Edge layer and Clients (end devices). The two levels of model aggregation are controlled by the "aggregation signal" sent from the cloud based on the information of local models collected by the edge nodes. In the system, clients are connected to the edge nodes via relatively low-speed (compared with high-speed network), shared (therefore noisy) wireless channels, whilst the edge-cloud connection is more stable (typically through Ethernet) and the bandwidth is typically sufficient.}  
\label{fig:overview}
\end{figure*}

Considering the privacy-preserving and communication-efficient nature of federated learning, it is now regarded as a promising approach to realizing intelligence on local devices and is a natural fit for the classic two-layer architectures such as cloud computing. However, with the emergence of MEC, there is still much room to explore in adapting the original FL protocol to the three-layer hierarchy of MEC, where abundant resources (in terms of computation, storage and network) are available in the edge layer. The necessity of such adaptation is two-fold. First, the network connection between the central server (i.e., the cloud) and the geographically distributed devices can be fairly slow and unreliable. Second, the central server may get overwhelmed by the workload or the network traffic due to excessive model updates from a vast number of clients. Introducing the edge layer, comprised of edge nodes such as base stations and servers in micro data centers, to the FL process can address these issues effectively. Edge nodes are proximate to the end devices and able to provide more stable connections and sufficient resources in computation. Existing work on edge-based FL \cite{IBM_FL}\cite{hierFAVG} has made use of edge resources or realized multi-step aggregation. However, the following issues have not been resolved yet: i) the \textbf{unreliability} of end devices (and their connections) is not considered; ii) the inefficiency of FL caused by the \textbf{heterogeneity} in end devices (and their bandwidth) is not addressed; iii) the \textbf{capability of edge nodes} (especially in terms of network) is not fully exploited.

The situation becomes even more challenging when combined with the strong privacy-preserving scenario, in which the servers are restricted from probing the information of clients \cite{FL_survey1}. For example, in the keyboard input prediction task, the device-server communication is stateless -- end devices can be invited to participate, but cannot be tracked, and their reliability (i.e., the chance that they drop/opt out during the training) is agnostic. 

In this paper, we propose a novel protocol (HybridFL) to enable privacy-preserving, efficient federated learning under the MEC architecture. Fig. 1 shows an overview of our protocol. We leverage the capability of edge nodes to boost the efficiency of communication and adapt FL to the three-layer hierarchy of MEC by making the FL process a hybrid of client-edge and edge-cloud collaborations. We also take into account both the unreliability and heterogeneity of end devices under strong privacy conditions and adopt a pace steering mechanism which is a hybrid of synchronous (edge-cloud) and asynchronous (client-edge) communication.. The main contributions of our work are outlined as follows:

\begin{itemize}
\item We propose a novel protocol (HybridFL) to drive the FL process in the three-layer architecture of MEC. HybridFL facilitates efficient model exchanges via the combination of quota-triggered regional aggregation (via the edge layer) and immediate cloud aggregation.
\item We mitigate the impact of client drop-out by introducing a regional slack factor for each edge node (i.e. region) into the client selection step via a probabilistic estimation method under strong privacy-preserving conditions that cause clients' reliability to be agnostic.
\item We introduce Effective Data Coverage (EDC) into the step of cloud-level model aggregation and present the convergence analysis for our protocol.
\item We conducted extensive experiments with machine learning tasks (using two public data sets) to evaluate the performance of our HybridFL protocol. Experimental results demonstrate significant improvement in average round duration, global model's convergence speed and accuracy, and the energy consumption of end devices.
\end{itemize}

The rest of this paper is organized as follows: Section II discusses the relevant studies concerning FL and MEC. In Section III, we detail the design of HybridFL and give convergence analysis for our protocol. In Section IV, we present and discuss the experimental results. We conclude this paper in Section V.

\section{Related Work}
\label{secII}
Stochastic Gradient Descent (SGD) \cite{SGD1}\cite{SGD2} and its variations have been the de facto standards for training most of the modern machine learning models. The fundamentals of previous distributed learning methods were built on the exchange of gradients, no matter in a centralized or decentralized manner. Extensive studies have shown the effectiveness of distributed SGD \cite{dist_SGD3}\cite{dist_SGD4} for complex models such as Deep Neural Networks (DNN). However, most of these traditional methods are designed for or tested in the data-center type of environments, where the data can be accessed globally by all workers and the financial cost of the communications is hardly considered as an issue. However, the real-life scenarios of MEC are usually comprised of low-spec, unreliable end devices, geographically distributed edge nodes with moderate performance, and relatively low-speed, noisy communication between edge nodes and end devices via wireless channels.

Federated Learning (FL), in a large part, addresses the problems of data privacy and prohibitive communication cost in training a global model from decentralized data. FL was originally designed as a synchronous training protocol called FedAvg \cite{FL}, which is a weight-averaging algorithm and aggregates the local models from end devices to produce a new global model each round. Model exchange, which is much less frequent (and thereby more communication-efficient) than the gradient exchange in traditional distributed SGD methods, is the outstanding feature of FL. Many Studies show that there is still potential in further reducing the communication cost of FL via model compression \cite{FL_comm}, setting adaptive aggregation intervals \cite{IBM_FL} and using multi-task learning \cite{MOCHA}. In addition, many variants of the FL protocol have also been proposed. For example, Xie et al. \cite{FedAsync} and Sprague et al. \cite{FedGeo} adopted asynchronous federated optimization schemes with non-blocking global model update and allowing devices to join halfway during training. SAFA \cite{SAFA} is semi-asynchronous FL protocol that retains the synchronized pace steering while introducing the strategies such as post-training client selection and model caching to speed up the training process.

It is natural to adapt FL to the emerging Mobile Edge Computing. A number of studies have provided their solutions to such adaptation. Wang et al. \cite{IBM_FL} took into account the resource budgets for edge nodes and proposed a pace control algorithm that adaptively adjusts the aggregation interval of FL. They adopted a system architecture in which the data reside on edge nodes, which essentially makes the system still a two-layer FL protocol. Liu et al. \cite{hierFAVG} implemented a hierarchical FL protocol that utilizes the cloud server and the edge nodes to perform two levels of model aggregation. The protocol is a straightforward extension of FedAvg, allowing multiple rounds in the edge layer before a global aggregation by the cloud. A main problem of the work is that each pair of interactions (i.e., client-edge and edge-cloud) is tightly coupled. As a result, device/network failures will cause both the edge nodes and the cloud to wait for a long time. In addition, in HierFAVG, edge nodes have to perform multiple rounds of edge-level aggregation before sending models to the cloud. This significantly postpones the global exchange of model information and consequently slows down the convergence. Considering the drawbacks of the existing solutions in driving FL in the MEC systems, in this work we aim to develop a more efficient FL protocol that enables fast, robust machine learning by virtue of the resources in the edge layer.

\section{The HybridFL Protocol}
\label{secIII}
There are eight steps in each round of training using the HybridFL protocol to drive FL in the MEC architecture (see Fig. \ref{fig:overview}). These eight steps form three basic stages: \textit{model distribution}, \textit{local training} and \textit{model aggregation}. The stage of model distribution starts with client selection (step 1, Fig. \ref{fig:overview}), after which the (latest) global model is distributed over the edge nodes (step 2) and then across all the clients (step 3). The second stage, local training, is performed on the clients (end devices) and also includes the steps of model downloading and uploading via the client-edge connections. Aggregation is the final stage of a round where the local models (selected and uploaded) are merged into a global model. Our protocol adopts a hybrid pace steering mechanism that allows flexible control over the edge-level (i.e, regional) model aggregation, which is signified by the cloud.%and then performs the cloud-level aggregation immediately after the edge-level aggregation to facilitate global model exchanges.

In this section we present the detailed design of HybridFL, in particular how we introduce regional slack factors into the model distribution stage and how our protocol performs the aggregations at the edge- and cloud-level. 

In this paper, we refer to the collection of clients connected to an edge node as a \textit{region}. $D^r$ denotes the set of data in region $r$ (note that the data cannot leave their end devices), i.e., $D^r = \{D^r_k | \forall k \in \mathrm{region\,} r \}$ . Without loss of generality, we assume a client can only connect to a single edge node. Table \ref{tab:symbols} lists the notations frequently used in this paper.

\begin{table}[ht]
\centering
\caption{List of symbols}
\begin{tabular}{ l l } 
 \hline
 Symbol 	  	& Description \\ 
 \hline
 $D$			& the complete dataset \\
 $D_k^r$		& the data partition on client $k$ in region $r$ \\
 $D^r$			& the (logical) set of data in region $r$ \\
 $n$			& the number of clients\\
 $n_r$			& the number of clients connected to edge node $r$\\
 $m$			& the number of edge nodes (regions)\\
 $V_C$			& the set of clients\\
 $V_E$			& the set of edge nodes\\
 $V_C^r$		& the set of clients in region $r$ ($|V_C^r|=n_r$)\\
 $w$			& parameters of the global model \\
 $w^r$			& parameters of the model on edge node $r$ \\
 $w_k^r$		& parameters of the local model on client $k$ \\
 $C$		 	& the desired proportion of clients that submit their local\\
 				& 	 models in a round. $C$ is specified by the cloud.\\
 $C_r$		 	& the proportion of clients selected in region $r$\\
 $U(t)$			& the set of selected clients in round $t$\\
 $U_r(t)$		& the set of selected clients within region $r$ in round $t$;\\
 				&	$|U_r(t)| = C_r \cdot n_r$\\
 $X(t)$			& the set of all clients (across all regions) that do not\\
 				& 	 drop out in round $t$\\
 $X_r(t)$		& the set of clients in $X(t)$ belonging to region $r$ \\
 $S(t)$			& the set of clients that submit their models in time and \\
 				&	 successfully in round $t$\\
 $S_r(t)$		& the set of clients in $S(t)$ belonging to region $r$\\
 \hline
\end{tabular}
\label{tab:symbols}
\end{table}

In the first stage of any FL round, client selection is often performed to ensure that only a reasonable proportion of clients are engaged in training this round. For example, the number of selected clients is determined by the proportion $C$ in \cite{FL}\cite{FL_sys}. As pointed out by Kairouz et al. \cite{FL_survey1}, it is necessary to restrict the participating population to a small fraction for two reasons. First, it has been shown that involving an excessive number of clients can hardly benefit the convergence and quality of the global model \cite{FL}. Second, recruiting excessive devices is neither cost-efficient in communication nor realistic for the end device owners.

Nevertheless, a severe shortage of participants in FL also leads to an inferior global model, of which the unreliability of end devices is the main cause. These devices can opt out of any round of local training or drop out occasionally due to various reasons such as low battery level, device failure or network disconnection. Let $X_r(t)$ denote the set of clients who are in region $r$ and do not drop out of round $t$. Then, since the client may drop-out, manually or unexpectedly, we have $|X_r(t)| \leq C\cdot n_r$ and $|X(t)| = \sum_{r=1}^m |X_r(t)| \leq C \cdot n$, where $C$ is the desired proportion of clients with successful model submission ($C$ is preset by the cloud server). 

In order to mitigate the shortage of participants caused by drop-out, we introduce $C_r(t)$ as the region-wise selection proportion into the client selection step (i.e., step 1 in Fig. \ref{fig:overview}) of the HybridFL training process at the start of each round. More specifically, an edge node $r$ will determine $C_r(t)$ and select a fraction of $C_r(t) \cdot n_r$ clients randomly (the set of selected clients is denoted by $U_r(t)$) before signifying these clients to begin local training in round $t$. An ideal value of $C_r(t)$ should satisfy that: i) the resulting $|U_r(t)|$ should be large enough so that the stragglers and dropouts have the minimal impact on round efficiency, and ii) $|U_r(t)|$ should not be too large, otherwise local training on some devices may be futile because the cloud only accepts a maximum of $C \cdot n$ clients each round. The main challenge here is that it is not permitted for an edge node to probe the state of its clients (including their IDs, aliveness, training progress, and the number of model updates made by a particular client), which causes the client's reliability (i.e., the probability that it drops out in a round) to be agnostic to the edge and the cloud. In view of this, we develop a probabilistic approach in this work to determine $C_r(t)$ for each region.

%In order to mitigate the shortage of participants caused by drop-out, we introduce a regional slack factor for each region into our protocol. The regional slack factor, denoted by $\theta_r(t)$, is applied to client selection for modulating the fraction of clients selected by each edge node $r$ in a FL round $t$. With $\theta_r(t)$, the proportion $C_r(t)$ of clients to be selected in region $r$ can be formulated as a function $g(\cdot)$:
%
%\begin{equation}
%	C_r(t) = g(C,\theta_r(t))
%\label{eq:C_r1}
%\end{equation}
%where $C$ is the desired update proportion of clients preset by the cloud server. The purpose of the regional slack factor $\theta_r(t)$ is to compensate the client dropouts. In this work, we do not need a priori knowledge regarding whether a client will drop-out in any round $t$. In the following section, we introduce how we determine the function $g(\cdot)$ and $\theta_r(t)$ using a probabilistic approach.

\subsection{Regional Client Selection}
By specifying the regional selection proportion $C_r(t)$ in round $t$, we aim to involve a fraction of $C_r(t)n_r$ clients in region $r$ and expect that $C\cdot n_r$ of them do not drop/opt out, provided that all these clients may be unreliable. Formally, the target of our region-wise selection can be formulated as:

\begin{equation}
	\mathbb{E}[|X_r(t)|; C_r^*(t), n_r] = C \cdot n_r
\label{eq:select_target}
\end{equation}
where $\mathbb{E}[|X_r(t)|; C_r^*(t), n_r]$ is the expectation of the number of clients in region $r$ that do not drop out in round $t$ given that an optimal proportion of clients $C_r(t)=C_r^*(t)$ are selected from $n_r$ (i.e., total number of clients in region $r$) clients to perform local training. 

Given any selection proportion $C_r(t)$, the expectation at the left-hand-side of (\ref{eq:select_target}) is equivalent to:

\begin{equation}
	\mathbb{E}[|X_r(t)|; C_r(t), n_r] = \sum_{k=0}^{C_r(t) n_r} k \sum_{b\in \text{comb}(U_r(t),k)} P(b)
\label{eq:EX2Pb}
\end{equation}
where $U_r(t)$ is the set of clients selected in region $r$, $\mathit{comb}(U,k)$ is the set of all combinations when selecting $k$ elements from the set of $U$, and $P(b)$ is the probability that the combination $b$ of end devices happen to be those who do not drop out in round $t$. Given a combination $b \in \mathit{comb}(U_r(t),k)$ and let $P_i^r(t)$ denote the probability that device $i$ of region $r$ does not drop/opt out in round $t$, $P(b)$ can be calculated by (\ref{eq:P(b)}):

\begin{equation}
	P(b) = \prod_{i \in b}P_i^r(t) \cdot \prod_{i \notin b}(1-P_i^r(t))
\label{eq:P(b)}
\end{equation}

Therefore, to obtain the optimal client selection proportion $C_r^*(t)$ we must solve it from (\ref{eq:optimal_Cr}) combining (\ref{eq:P(b)}):

\begin{equation}
	\sum_{k=0}^{C_r^*(t) n_r} k \sum_{b\in \text{comb}(U_r(t),k)} P(b) = C \cdot n_r
\label{eq:optimal_Cr}
\end{equation}

However, $C_r^*(t)$ cannot be solved from (\ref{eq:optimal_Cr}) without a priori knowledge on the probability $P_i^r(t)$ (i.e., reliability) of every individual client. In this work we consider a FL scenario with strong privacy-preserving condition, under which it is prohibited to acquire the clients' identifiers and their states \cite{FL_survey1}, i.e., $P_i^r(t)$ is agnostic. In view of this, we develop a novel approach to address this difficulty and eventually work out the optimal value of $C_r(t)$ for each region $r$ each round $t$. 

Assume that $\theta_r(t)$ is such a probability that after we replace each individual $P_i^r(t)$ in (\ref{eq:P(b)}) with $\theta_r(t)$, the resulting expectation of $|X_r(t)|$ remains unchanged. We can always find such $\theta_r(t)$ because after the replacement, $|X_r(t)|$ of region $r$ follows the Binomial distribution $\mathcal{B}(C_r(t) n_r, \theta_r(t))$ and the expectation of $|X_r(t)|$ (i.e., $\mathbb{E}[|X_r(t)|; C_r(t), n_r] \in [0, C_r(t)n_r]$) is a surjective function of $\theta_r(t) \in [0,1]$. %which is an estimation of the no-abort probability given any client performing training in region $r$. By doing so we screen the discrepancy (in reliability) of clients in the same region but retain that inter-regionally.
Now we can re-write the right-hand side of (\ref{eq:EX2Pb}):

\begin{align}
	\mathbb{E}[|X_r(t)|; C_r(t), n_r]  	&=  \sum_{k=0}^{C_r(t) n_r} k \cdot P(|X_r(t)| = k) \nonumber \\
										&=  C_r(t) n_r \theta_r(t)
\label{eq:EX2theta}
\end{align}
where the second equality in (\ref{eq:EX2theta}) holds because $|X_r(t)| \thicksim \mathcal{B}(C_r(t) n_r, \theta_r(t))$. 

Combining (\ref{eq:EX2theta}) and our selection target (\ref{eq:select_target}), we have:

\begin{equation}
	C_r(t) = \frac{C}{\theta_r(t)} \;\; \text{if } C_r(t) = C_r^*(t)
\label{eq:Cr2theta}
\end{equation}
where $C$ is the desired global proportion of clients (specified by the cloud) with successful model submissions in round $t$ over the entire MEC system. $\theta_r(t)$ defined in (\ref{eq:EX2theta}) modulates the selection proportion in a region to compensate the client drop-out in that region. Therefore we term $\theta_r(t)$ the \textit{regional slack factor} (for region $r$). 

Note that $\theta_r(t)$ in (\ref{eq:Cr2theta}) cannot be decided arbitrarily, otherwise the optimality of $C_r(t)$ is not guaranteed. This is because there is only one optimal value for $C_r(t)$ given any distribution of client reliability and the target formulated in (\ref{eq:select_target}). In other words, if we determine $C_r(t)$ via (\ref{eq:Cr2theta}) provided an under-estimated $\theta_r(t)$, the selection proportion $C_r(t)$ will be too big (i.e., $C_r(t) > C_r^*(t)$) for region $r$ and consequently, the target expectation of $|X_r(t)|$ will be higher than the desired level, i.e., $\mathbb{E}[|X_r(t)|; C_r(t), n_r] > C \cdot n_r$. It is similar for the situation of over-estimated $\theta_r(t)$.

%This is because (\ref{eq:EX2theta}) is derived from (\ref{eq:EX2Pb}) and the formulation of (\ref{eq:EX2Pb}) shows that it is an one-to-one mapping between $C_r(t)$ and the expectation $\mathbb{E}[|X_r(t)|; C_r(t), n_r]$. Therefore, in order to achieve the target specified in (\ref{eq:select_target}), there is only one optimal value for $C_r(t)$ given any distribution of client reliability. In other words, if we determine $C_r(t)$ via (\ref{eq:Cr2theta}) provided an under-estimated $\theta_r(t)$, the selection proportion $C_r(t)$ will be too big (i.e., $C_r(t) > C_r^*(t)$) for region $r$ and consequently, the target expectation of $|X_r(t)|$ will go higher than the desired level, i.e., $\mathbb{E}[|X_r(t)|; C_r(t), n_r] > C \cdot n_r$. Vice versa for the situation of over-estimated $\theta_r(t)$.

According to (\ref{eq:Cr2theta}), we can determine how many clients we need to select in each region for a upcoming FL round after $\theta_r(t)$ is resolved. In this work, we develop a novel method to estimate $\theta_r(t)$ based on the historical records of model submissions (since the course of FL is organized in rounds), i.e., how many models are collected by each region in previous rounds. Note that edge nodes can only count the models they collected but do not know which client submitted the model. 

In HybridFL, we adopt a quota-triggered aggregation mechanism in which the cloud ends a round once $C\cdot n$ client models have been submitted globally across the MEC system. As a result, we have:

\begin{equation}
	\sum_{r\in V_E}|S_r(t)| = \min(nC, \,\sum_{r\in V_E}|X_r(t)|)
\label{eq:SandX}
\end{equation}
where $S_r(t)$ is the set of clients that submit their models in time in round $t$ (before the cloud ends a round after collecting $C\cdot n$ models globally) and $V_E$ is the set of edge nodes. Note that $S_r(t) \subseteq X_r(t)$ because when the cloud ends a round, some clients may be still working and have not finished local training. Details of how to determine the aggregation timing will be introduced later. Formally, we use a factor $q_r^*(t)$ to characterize the relation between $|S_r(t)|$ and $|X_r(t)|$:

\begin{equation}
	|S_r(t)| = |X_r(t)| \cdot q_r^*(t)
\label{eq:pct_star}
\end{equation}
where $q_r^*(t)$ denotes the percentage of clients in $X_r(t)$ that submit local models in time (these clients make up $S_r(t)$). Note that $|S_r(t)|$ is observable as the number of local models collected by edge node $r$ in round $t$. However, $X_r(t)$ is agnostic since we consider a strong privacy-protection scenario where the edge nodes are not allowed to probe the state of clients. We can only observe how many clients submitted the updated models (i.e., $|S_r(t)|$) but cannot know who have dropped out and who are still working. Therefore, we transform (\ref{eq:pct_star}) into (\ref{eq:pct_star2}) given $\mathbb{E}[|X_r(t)|; C_r(t), n_r] \neq 0$, and then define $q_r(t)$ in (\ref{eq:pct}).

\begin{equation}
	|S_r(t)| = \mathbb{E}[|X_r(t)|; C_r(t), n_r] \cdot \frac{|X_r(t)| \cdot q_r^*(t)}{\mathbb{E}[|X_r(t)|; C_r(t), n_r]}
\label{eq:pct_star2}
\end{equation}

\begin{equation}
	q_r(t)	\triangleq \frac{|X_r(t)| \cdot q_r^*(t)}{\mathbb{E}[|X_r(t)|; C_r(t), n_r]} 
\label{eq:pct}
\end{equation}

From (\ref{eq:pct_star2}), (\ref{eq:pct}) and (\ref{eq:EX2theta}), we have:

\begin{align}
	|S_r(t)|	& = \mathbb{E}[|X_r(t)|; C_r(t), n_r] \cdot q_r(t)  \nonumber \\
				& = C_r(t) n_r \theta_r(t) \cdot q_r(t)
\label{eq:recur1}
\end{align}

Note that the value of $\theta_r(t)$ needs to be estimated before round $t$ starts so that we can determine the selection proportion $C_r(t)$ (every round begins with the client selection step). However, (\ref{eq:recur1}) cannot be used directly to obtain $\theta_r(t)$ because $S_r(t)$ and $q_r(t)$ are unknown before round $t$ is completed (Note that $S_r(t)$ is observable at the end of round $t$ so we can calculate $q_r(t)$ with $|S_r(t)|$ at round ends by (\ref{eq:pct-2}) combining (\ref{eq:pct}), (\ref{eq:pct_star}) and (\ref{eq:select_target}) with the assumption that $C_r(t)$ is the optimal).
%Note that we can calculate $q_r(t)$ with $S_r(t)$ at round end by (\ref{eq:pct-2}) combining (\ref{eq:pct}), (\ref{eq:pct_star}) and (\ref{eq:select_target}) with the assumption that $C_r(t)$ is the optimal. 

\begin{equation}
	q_r(t)	= \frac{|S_r(t)|}{C \cdot n_r} 
\label{eq:pct-2}
\end{equation}

Therefore we develop the following practical approach to work out $\theta_r(t)$ by exploiting the historical records of the variables observable to edge nodes. More specifically, edge node $r$ has stored $S_r(1), S_r(2), \ldots, S_r(t-1)$, $q_r(1), q_r(2), \ldots, q_r(t-1)$ and $C_r(1), C_r(2), \ldots, C_r(t-1)$ at the start of round $t$. Also, according to the definition of $\theta_r(t)$, it represents a region-wise property. So we assume $\theta_r(t)$ does not change significantly over the course of the FL training. Thus, we use a constant $\hat{\theta}_r(T)$ as the approximation of $\theta_r(t)$ within the time window $T$ spanning from round 1 to round $t$:

\begin{equation}
	\theta_r(i) \approx \hat{\theta}_r(T),\, \forall i \in \{1,2,\ldots,t\}
\label{eq:theta_hat}
\end{equation}

Replacing $\theta_r(i)$ with $\hat{\theta}_r(T)$ in (\ref{eq:recur1}) and for round $i$, $\forall i<t$, we have: 

\begin{equation}
	\frac{|S_r(i)|}{n_r} \approx C_r(i) q_r(i) \hat{\theta}_r(T),\, \forall i \in \{1,2,\ldots,t-1\}
\label{eq:recur2}
\end{equation}

Therefore, (\ref{eq:recur2}) is equivalent to a series of observations (the number of which is $t-1$) sampled from a function in the form of "$y = ax$" (i.e., $|S_r(i)|/n_r$ and $C_r(i)q_r(i)$ being the samples of $y$ and $x$, respectively, and $\hat{\theta}_r(T)$ is the coefficient). In view of this, we use Least Square Estimation (LSE) to fit the value of $\hat{\theta}_r(T)$ based on (\ref{eq:recur2}), which produces:

\begin{equation}
	\hat{\theta}_r(T) \stackrel{LSE}{=} 
		\frac{1}{n_r} \frac{\sum_{i=1}^{t-1} C_r(i)q_r(i) |S_r(i)|}{\sum_{i=1}^{t-1} \big(C_r(i)q_r(i)\big)^2}
		,\; t>1.
\label{eq:LSE}
\end{equation}
where $S_r(i)$, $C_r(i)$ and $q_r(i), i=1,2,\ldots,t-1$ are retrieved from the logs of edge node $r$. At the start of round $t$, we compute $\hat{\theta}_r(T)$ and use it as an estimate of $\theta_r(t)$, and then determine region $r$'s client selection proportion $C_r(t)$ (defined in (\ref{eq:Cr2theta})) using (\ref{eq:C_r_final}):

\begin{equation}
	C_r(t) = C\cdot n_r \frac{\sum_{i=1}^{t-1} \big(C_r(i)q_r(i)\big)^2}{\sum_{i=1}^{t-1} C_r(i)q_r(i) |S_r(i)|},\; t>1.
\label{eq:C_r_final}
\end{equation}

For $t$=1 (the 1st round of FL), $\theta_r(t)$ is initialized as a default value (e.g., $\theta_r(1)$=0.5). $C_r(1)$ is initialized to $C/\theta_r(1)$ accordingly. To investigate the effectiveness of our method in terms of achieving the selection target (\ref{eq:select_target}), we simulated 20 clients in two regions and ran 100 rounds (5 local epochs in each round) of federated learning using HybridFL as the control protocol. We initialized $\theta_r(1)$ to 0.5.

% tracing
\begin{figure}[htb]
    \centering
    \includegraphics[width=250px]{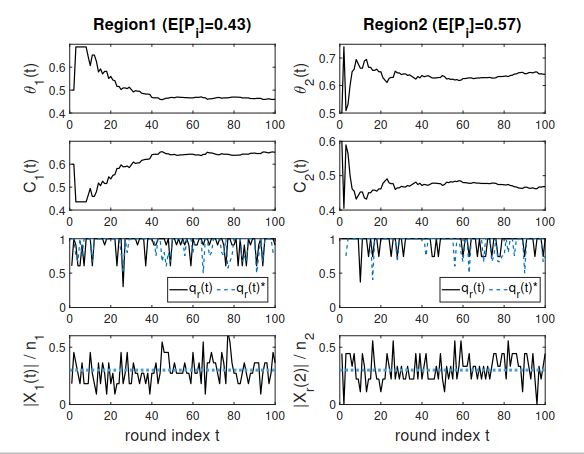}
    \caption{The traces of $\theta_r(t)$, $C_r(t)$, $q_r(t)$ and $|X_r(t)|/n_r$ in a simulation where there are 11 and 9 clients in Region 1 and Region 2, respectively. The reliability of clients in training follows the Gaussian distribution $\mathcal{N}(\mu, 0.15^2)$ where $\mu=\mathbb{E}[P_i^r(t)]$ is set to 0.43 and 0.57 for Region 1 and Region 2, respectively. Clients also differ in performance which follows $\mathcal{N}(0.5, 0.1^2)$; The global selection fraction $C$ is set to $0.3$.}
\label{fig:trace_theta}
\end{figure}

From the traces of $\theta_r(t)$, $C_r(t)$, $q_r(t)$ and $|X_r(t)|/n_r$ in Fig. \ref{fig:trace_theta}, we can observe that our probabilistic estimation drives $\theta_r(t)$ and $C_r(t)$ (the first two rows in the figure) to the convergence at about 40 rounds of FL. Note that $\theta_1(t)$ and $\theta_2(t)$ converge to 0.46 and 0.63, which, by the definition, are not necessarily equal to $\mathbb{E}[P_i^r(t)]$ (recall that $P_i^r(t)$ is the reliability of client $i$ in region $r$) which is set to 0.43 and 0.57 for Region 1 and Region 2 in this example, respectively. Besides, we define $q_r(t)$ without using any knowledge about $X_r(t)$, but still produces a close approximation to its true value $q_r^*(t)=|S_r(t)|/|X_r(t)|$ (the 3rd row in Fig. \ref{fig:trace_theta}). Consequently, the client participating ratio in a region, quantified by $|X_r(t)|/n_r$, is maintained around $C=0.3$ (shown in the last row of Fig. \ref{fig:trace_theta}; the blue dash line represents $C=0.3$) after the convergence of $\theta_r(t)$ and $C_r(t)$. 

With this case we demonstrate that our method for estimating $\theta_r(t)$ (which determines $C_r(t)$) is both theoretically and practically feasible for finding the optimal value of the regional selection proportion that leads to the very expectation of $|X_r(t)|$ desired by the cloud (see the target (\ref{eq:select_target})).

\subsection{Model Aggregations}
In our protocol (HybridFL), model aggregation is a multi-step stage, involving both edge- and cloud-level aggregation (see steps 6, 7, 8 and 9 in fig. \ref{fig:overview}). In HybridFL, once the updated models submitted by the clients across the MEC system equals to $C\cdot n$ (i.e., $|S(t)|=\sum_{r\in V_E} |S_r(t)|$ reaches $C\cdot n$), it triggers the cloud to send the "aggregation signal" to the edge nodes (see step 6, Fig. \ref{fig:overview}). The edge nodes will then stop waiting for more local models. This quota-triggered regional aggregation effectively mitigates the impact of the clients which have dropped out or are straggling. Consequently the round length is expected to be shortened (our experiment results support this expectation).  

\begin{figure}[htb]
    \centering
    \includegraphics[scale=0.43]{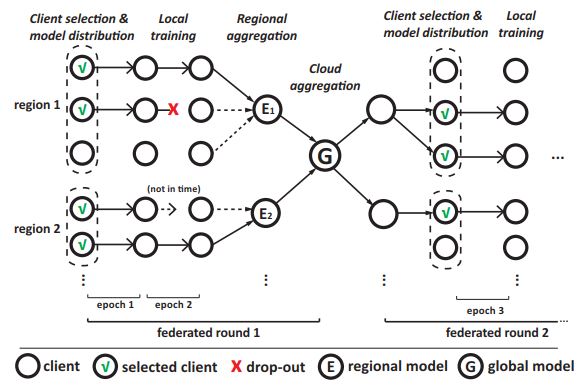}
    \caption{The workflow of the proposed HybridFL protocol wherein the cloud requires two local model submissions (since $C$=0.4 and $n$=5 in this example) each round to trigger the model aggregation. Local training consists of two epochs. Note that the models from the dropped-out or straggling clients (e.g., clients no. 2 and 4 in round 1) are not actually uploaded (dashed line with arrow).}
    \label{fig:workflow}
\end{figure}

We adopt an \textit{immediate cloud aggregation} strategy, which allows the cloud-level model aggregation to be conducted right after the edge-level aggregation is completed. The rationale behind this strategy is that the cloud-edge network connection is typically  reliable and of low latency. Therefore, it facilitates the global information exchange and the convergence of the global model by aggregating the regional models at the cloud level as early as possible after the regional aggregations are completed at the edge nodes. Fig. \ref{fig:workflow} demonstrates how rounds are orchestrated in HybridFL.

% FL workflow

The cloud keeps monitoring the total number of clients that have submitted their models each round by listening to the reports of the current value of $|S_r(t)|$ from the edge nodes. Once the total number of client model submissions reaches the quota $C\cdot n$, the cloud will signify the edge nodes to perform regional aggregation, the result of which can be formulated as:

\begin{equation}
	w^r(t) = \sum_{k \in V_C^r} \frac{|D_k^r|}{|D^r|} w_k^r(t)
\label{eq:w_r}
\end{equation}
where $w_k^r(t)$ is the model on client $k$ in region $r$ and $w^r(t)$ denotes the resulting regional model for edge node $r$ in round $t$. Note that the aggregation involves all client models in the region, not limited to those who finished local training successfully (see fig. \ref{fig:workflow}). To alleviate model staleness, we use a cache solution in which the local models without successful update in the current round are replaced with the existing regional model obtained in last round before the aggregation is conducted, i.e.,  $w_k^r(t) = w^r(t-1)$ if $k \notin S_r(t)$.

The cloud aggregation will be performed immediately after the regional aggregation to produce the cloud model. Instead of using constant weight for each regional model as in the literature \cite{hierFAVG}, we adopt a data-oriented weight averaging strategy by introducing the \textit{Effective Data Coverage} (EDC) for each region in every round. EDC quantifies the actual size of data covered in round $t$'s training based on $S_r(t)$. We formulate EDC for region $r$ in round $t$ (denoted by $EDC_r(t)$) as:

\begin{equation}
	EDC_r(t) = \sum_{k \in S_r(t)} |D_k^r|
\label{eq:EDC_r}
\end{equation}
where $S_r(t)$ is the set of clients who submitted their models successfully to its regional edge node. Accordingly, we further define EDC for the whole MEC system (denoted by $EDC(t)$) as:

\begin{equation}
	EDC(t) = \sum_{r \in V_E} EDC_r(t).
\label{eq:EDC}
\end{equation}

In the model aggregation step at the cloud level, we weight each regional model $w^r(t) $ based on EDC to characterize its round-wise contribution in producing the global model $w(t)$:

\begin{equation}
	w(t) = \sum_{r \in V_E} \frac{EDC_r(t)}{EDC(t)} w^r(t).
\label{eq:w}
\end{equation}

Algorithm \ref{algo} presents the pseudo-code of the entire process of FL using our protocol.

% Algorithm, HybridFL
\begin{algorithm}[htb] 
\caption{the HybridFL protocol}
	\DontPrintSemicolon
	\SetKwInOut{Input}{Input}
	\SetKwInOut{Output}{Output}
	\SetKwProg{Pn}{}{:}{\KwRet}  % programme
	\SetAlgoLined
	\Input{maximum number of rounds $t_{max}$, local epochs per round $\tau$, desired proportion $C$, response time limit $T_{lim}$}
	\Output{finalized global model $w$}
	% use \rm before each statement containing text	
	\smallskip
  
  	\rm// Cloud process: running on the central server\;
    	\rm Initializes global model $w(0)$\;
    	$quota \leftarrow C \cdot n$\;
  		\For{\rm round $t\leftarrow 1$ to $t_{max}$}{
      		Distributes $w(t-1)$ to all the edge nodes\;
      		\For{\rm each edge node $r$ in $V_E$ \bf in parallel}{
       			Computes $C_r(t)$ according to (\ref{eq:C_r_final})\;
      			\it{edgeUpdate}($r, C_r(t), \tau$)\;
      	}
      	\rm Keeps monitoring update count by edge nodes\;
      	\If{\rm $|S(t)| \geq quota$ or $T_{lim}$ is reached}{
      		// \rm triggers regional aggregation\;
      		Sends aggregation signal to all edge nodes\;
			\it{edgeAggregation}($r$)\;
  	  	}
      	// \rm cloud aggregation\;
      	\rm Computes $w(t)$ according to (\ref{eq:w})\;
  	} % end For
  	\rm return $w(t)$\;

	\smallskip
	\rm// Edge process: running on edge node $r$\;
	\Pn{\bf edgeUpdate($r,C_r,\tau$)}{
        $q \leftarrow C_r \cdot n_r$\;
  		\rm $Q \leftarrow q$ randomly selected clients in region $r$\;
  		\For{\rm each client $k$ in $Q$ \bf in parallel}{
  			\it clientUpdate($k,r,\tau$)\;
  			\rm Keeps reporting update count to the cloud\;
  		}
  	}
  	\Pn{\bf edgeAggregation($r$)}{
  		\rm Computes $w^r(t)$ according to (\ref{eq:w_r})\;
  	}
  	
  	\smallskip
  	\rm// Client process: running on client $k$\;
  	\Pn{\bf clientUpdate($k,r,\tau$)}{
  		\For{\rm epoch $e\leftarrow 1$ to $\tau$}{
    		\rm Updates $w^r_k(t)$ using Gradient Descent method\;
  		}
  	}
  
\label{algo}
\end{algorithm}

\subsection{Convergence Analysis}
The convergence of the global model by federated learning has been proved for both the two-layer architecture \cite{IBM_FL} and the three-layer edge computing systems \cite{hierFAVG}. However, since we have made modification to the aggregation rules, it is necessary to provide the convergence analysis with the focus on showing the difference compared to the proof provided in the existing work.

Since the regional aggregation in HybridFL is followed instantly by the cloud (global) aggregation, we can mathematically re-formulate the global model $w(t)$ by combining (\ref{eq:w_r}) and (\ref{eq:w}), which yields (\ref{eq:reform-wt}):

\begin{align}
	w(t)& = \sum_{r\in V_E}\frac{EDC_r(t)}{EDC(t)} \sum_{k \in V_C^r} \frac{|D_k^r|}{|D^r|} w_k^r(t) \nonumber \\
		& = \sum_{k \in V_C} \frac{EDC_{r(k)}(t)}{EDC(t)} \frac{|D_k^{r(k)}|}{|D^{r(k)}|} w_k^{r(k)}(t) \nonumber \\
		& \triangleq \sum_{k \in V_C} \gamma(k,r(k),t) \cdot w_k^{r(k)}(t)
\label{eq:reform-wt}
\end{align}
where $r(k)$ stands for the corresponding edge node connected to client $k$ and the symbol $\gamma(k,r(k),t)$ represents the weight of client $k$'s model during the aggregation. Without ambiguity, in the rest of this paper, we use the abbreviation $\gamma(k,r,t)$ to denote $\gamma(k,r(k),t)$ for brevity. 

In (\ref{eq:reform-wt}), the first equality represents the two-level aggregation, namely, the second "$\sum$" represents the edge-level aggregation while the first "$\sum$" represents the cloud-level aggregation. The second equality in (\ref{eq:reform-wt}) transforms the two "$\sum$" into one. This suggests that the entire aggregation in our three-layer MEC system is equivalent to the two-layer FL process as shown in \cite{FL} with the only difference lying in weights.

Formally, the HybridFL process is to solve the following global optimization problem at any given round $t$:

% optimziation target
\begin{equation}
    \arg\min_{w} F(w,t) = \sum_{k \in V_C} \gamma(k,r,t) F_k(w)
\label{eq:target_F}
\end{equation}
where $w$ denotes the parameters (e.g., weights for neural nets) of the global model to be optimized, $r=r(k)$ is the region index for client $k$, and $F_k(w)$ is the average loss from the data partition on client $k$. $F_k(w)$ is calculated by:

\begin{equation}
    F_k(w) = \frac{1}{|D_k^{r(k)}|} \sum_{(x_i,y_i) \in D_k^{r(k)}} f(w; x_i,y_i)
\label{eq:F_kr}
\end{equation}
where $f(\cdot)$ is the loss function and $D_k^{r(k)}$ is the data possessed by client $k$ with $r(k)$ being its region. %The optimization objective is in accordance with that in \cite{FL} and \cite{IBM_FL} when there is no dropout and $C$=1 (which makes $\gamma(k,r,t)\equiv |D^r_k|/|D|$). %For clarity, in the rest of this paper we refer to a client by specifying both its index and the index of the edge node it connects to. 

We analyze the convergence of our protocol by quantifying the upper bound of $F(w(t),t)-F(w^*,t)$, where $w^*$ denotes the optimal model parameters for our target (see (\ref{eq:target_F})). Due to the space limit, we base our proof on the analysis provided by Wang et al. \cite{IBM_FL} and extend their \textit{Theorems 1 and 2} for the case of our protocol, which yields the \textbf{Theorem} \ref{theorem1} and \textbf{Theorem} \ref{theorem2} in our paper, respectively. We first make the following assumption to facilitate the analysis: 

% Assumption
\begin{assump}[Loss function]
	\textit{$F_k(w)$ is convex, $\rho$-Lipschitz and $\beta$-smooth.}
\label{assump}
\end{assump}

For the loss functions that do not satisfy the assumption above, Wang et al. \cite{IBM_FL} still validated the effectiveness of FL in such cases. With the assumption, we have: \textit{$F(w,t)$ is convex, $\rho$-Lipschitz and $\beta$-smooth} with regards to $w$, which can be proved using the triangle inequality based on (\ref{eq:target_F}). We also define $\delta_k$ as the upper bound of the divergence between the gradients of $F_k(w)$ and $F(w,t)$, and $\bar{\delta}$ as the upper limit of $\delta_k, \forall k \in V_C$:

\begin{equation}
	\|\nabla F_k(w) - \nabla F(w,t)\| \leq \delta_k \leq \bar{\delta}.
\label{eq:delta_k}
\end{equation}

Let $z$ denote the index of epoch (i.e., $z = 1,2,\ldots,t_{max} \cdot \tau$) ($\tau$ is the number of local epochs in a round). To facilitate the analysis, by $w([z])$ we denote a hypothetical global model as the result of aggregating all $w_k^r([z])$ at epoch $z$. It is not to be confused with $w(t)$ because $w(t)$ is only visible after the aggregation at the end of a round. Besides, we also consider an auxiliary model $v_t([z])$ learned using \textit{centralized} gradient descent initialized as $w(t-1)$ in the context of round $t$ for optimizing the same target $F(w,t)$. Given $z$ as an epoch in round $t$ (i.e., $z \in ((t-1)\tau, t\tau]$), by the definitions we have:

\begin{align}
	w([z]) 	& = \sum_{k \in V_C} \gamma(k,r,t) w_k^r([z])
\label{eq:w(z)_update}
\end{align}
where $w_k^r([z])$ is updated from $w_k^r([z-1])$:
\begin{align}
	w_k^r([z]) = w_k^r([z-1]) - \eta \nabla F_k(w_k^r([z-1]))
\label{eq:w_k(z)_update}
\end{align}

For $v_t([z])$ with $z \in ((t-1)\tau, t\tau]$, we have:
\begin{equation}
	v_t([z]) = v_t([z-1]) - \eta \nabla F(v_t([z-1]), t)
\label{eq:v(z)_update}
\end{equation}

Now we give our theorem \ref{theorem1}:

\begin{customthm}{1*}[Loss divergence bound]
	\it for any epoch $z$ in round $t$, we have
	\begin{equation}
		F(w([z]),t) - F(v_t([z]),t) \leq \rho \bar{h}(z-(t-1)\tau)
	\end{equation}
	\it where 
	\begin{equation}
		\bar{h}(x) \triangleq \frac{\bar{\delta}}{\beta} ((\eta \beta + 1)^x -1) - \eta \bar{\delta}x
	\label{eq:bar_h}
	\end{equation}
\label{theorem1}
\end{customthm}

\noindent \textit{Proof.} We base our proof of Theorem \ref{theorem1} on [Lemma 2, ref. \cite{IBM_FL}]. See Appendix \ref{apdx:prf} in this paper for proof details. 
	
Theorem \ref{theorem1} gives the theoretical difference in loss between the global model $w([z])$ (by aggregating local models) and the baseline $v_t([z])$ (learned on centralized data) during the training process in round $t$. Note that $F(w([z]),t) - F(v_t([z]),t) \leq \rho\bar{h}(\tau)$ at $z=t\tau$ and $\bar{h}(1)=0$. This means that $w([z])$ is equivalent to $v_t([z])$ if the aggregation interval $\tau=1$. Based on Theorem \ref{theorem1} and recalling that $w(t)=w([t\cdot \tau])$, we now present the convergence upper bound of $w(t)$ in Theorem \ref{theorem2}:

\begin{customthm}{2*}[Convergence upper bound]
	\it After $t$ rounds with $\tau$ epochs in each round, the convergence of the global model is guaranteed by:
	\begin{equation}
		F(w(t),t) - F(w^*,t) 
			\leq \frac{1}{t\tau \big(\omega \eta (1-\frac{\beta \eta}{2})- \frac{\rho \bar{h}(\tau)}{\tau \epsilon^2} \big)}
	\end{equation}\\
	\it when the conditions below are satisfied:\\
	\it 1) $\eta \leq \frac{1}{\beta}$\\
	\it 2) $\omega \eta (1-\frac{\beta \eta}{2})- \frac{\rho \bar{h}(\tau)}{\tau \epsilon^2} > 0$\\
	\it 3) $F(v_t([z]),t) - F(w^*,t) \geq \epsilon,\, \forall z \in ((t-1)\tau, t\tau]$\\
	\it 4) $F(w(t),t) - F(w^*,t) \geq \epsilon$\\
	\it where $\epsilon > 0$, $\omega \triangleq \mathrm{min}_t \frac{1}{\|v_t([(t-1)\tau])-w^*\|^2}$, and $\bar{h}(\cdot)$ is defined in (\ref{eq:bar_h}).
\label{theorem2}
\end{customthm}

\noindent \textit{proof.} Condition 1 places a limit on the learning rate $\eta$ whilst condition 2 implies that the gap between $w([z])$ and $v_t([z])$ needs to be small. Conditions 3 and 4 limit the lower bound of the gap to a positive value $\epsilon$ because $w(t)$ is an approximation of $w^*$ given that we perform the aggregation after every $\tau$ local epochs and $\tau > 1$. Theorem \ref{theorem2} can be proved based on the conclusion of Theorem \ref{theorem1} combined with the [Lemmas 1, 3 and 4, ref. \cite{IBM_FL}], and the steps of proof are the same as that provided in \cite{IBM_FL}.  

Theorem \ref{theorem2} implies that the gap between $w(t)$ and $w^*$ (the optimum) in terms of optimizing the target loss function (see (\ref{eq:target_F})) narrows as the FL process proceeds, i.e., $t$ increases.

\subsection{Client Heterogeneity}
The heterogeneity of end devices is a common property of practical MEC systems. FL in such systems can involve a vast number of heterogeneous end devices, whose discrepancy in capability (e.g., CPU performance, bandwidth) and reliability has the major impact on the overall efficiency of FL. In this paper, we characterize the heterogeneity of clients by mainly considering their compute performance, bandwidth and reliability. 

The compute performance of a client determines how efficiently it conducts local training and can be measured by the CPU frequency (in GHz). Given the same training task and the same size of data partition, clients with lower performance require more time for local training. A certain space of memory is required for any on-device training process. For simplicity we assume that clients only participate when they have sufficient memory, and that memory does not impact clients' performance. After on-device training is completed, local models need to be transmitted to the edge nodes. In this step, device bandwidth is the main factor that determines the communication time between the edge and end devices. Due to the heterogeneity of clients, the time needed for model download/upload and local training differs from client to client. The heterogeneity of clients also differentiates them in energy consumption, which is determined jointly by their power consumption in computation and communication and the time needed for local training (computation) and model transmission (communication). The formulation of device performance and energy consumption is detailed in Section \ref{secIV}.

We also assume that clients are discrepant in reliability because they may drop/opt out by a different probability. Practically, the causes of client drop-out can involve many factors including any subjective/objective reasons, and they all vary from situation to situation. The correlation between these factors and how likely a client drops out is very complicated and beyond the scope of this work. In this paper, we consider client drop-out as an independent event. Our protocol is designed to be completely independent on the distribution of clients' drop-out probability (i.e., designed for reliability-agnostic scenarios).

The design of HybridFL mitigates the negative impact of end devices' heterogeneity and unreliability on the efficiency and effectiveness of FL. The introduction of regional slack factors enables our protocol to modulate the number of model submissions for each region based on the desired proportion controlled by the cloud. The slack factors are determined without any a priori knowledge on any client's drop-out probability. Besides, the quota-triggered aggregation mechanism in HybridFL allows the cloud to end a round once the quota is met, rather than passively awaiting response from every selected client. This effectively accelerates an FL round and makes the protocol less susceptible to device failure.

\section{Experimental Evaluation}
\label{secIV}
We evaluated the effectiveness of the proposed HybridFL in terms of model convergence speed, round efficiency and the global model's accuracy. We also evaluated the energy consumption of end devices, which we consider as an important metric in practice.

\subsection{Experiment Setup}
In the evaluation, we built a simulated MEC system for Federated Learning as a complete software package. The MEC system was established with simulated parties (i.e., the cloud, edge nodes and end devices) that comprise the three-layer architecture. On-device training in the FL process was implemented using the PyTorch framework. Each group of end devices (clients) are managed by and connected to an edge node via wireless channels, which forms a region, whilst the edge nodes and the cloud are connected through high-speed Ethernet. All the clients and their local network are implemented as being unreliable in the simulated MEC system. The drop-out probability of client $k$ is set as $dr_k$, which follows a Gaussian distribution (see Table \ref{tab:exp_setup}) with its mean value set to $\mathbb{E}[dr]$. For client $k$, the relation between its no-abort probability $P_k$ and its drop-out probability is: $P_k = 1 - dr_k$.

We evaluated HybridFL in two machine learning tasks: \textit{Aerofoil} (Task 1) and \textit{MNIST} (Task 2). For the tasks we configured different MEC environments to test the performance of HybridFL with different scale of end devices and edge nodes on different data distribution. The size of data partitions in each end device in Task 1 follows the Gaussian distribution while in Task 2, we set it to be non-IID by assigning the samples of class $y_i$, by a probability of 0.75, to the clients with indices $k\equiv y_i$ (mod 10). 

We also implemented two existing protocols in recent literature: \textit{FedAvg} \cite{FL} and \textit{HierFAVG} \cite{hierFAVG}. FedAvg is the primitive FL protocol proposed by Google for the two-layer client/server architecture. HierFAVG is a three-layer FL protocol for Edge Computing systems and adopts a similar training architecture as our protocol by introducing the edge layer that performs the edge-level model aggregation before the global aggregation conducted by the cloud. HierFAVG has no adaptive control over the flow of models. Both the edge and the cloud have to await the responses from all the selected clients. We compare the FL training process driven by these protocols and our HybridFL under the same settings. The parameters in our experimental setting are listed in Table \ref{tab:exp_setup}.

%[table: setup]
\begin{table}[ht]
\centering
\caption{Experimental setup for federated learning and the MEC system. The units of performance, bandwidth and throughput in the table are GHz, MHz, and Mbps, respectively.}
\begin{tabular}{ l l l l } 
 \hline
 Setting	 	  			& Symbol		&Task 1 					&Task 2  	\\ 
 \hline
 dataset		 			& $D$			&Aerofoil					&MNIST		\\
 \# of features				& $d$			&5							&28x28		\\
 model						& $w$			&FCN						&LeNet-5	\\
 dataset size	 			& $|D|$			&1503						&70k	   \\
 \# of clients	 			& $n$			&15							&500	   \\
 \# of edge nodes 			& $m$			&3							&10		   \\
 data distribution${}^1$	& -	  			&$\mathcal{N}(100,30^2)$	&non-IID, 0.75\\
 client performance			& $s_k$			&$\mathcal{N}(0.5,0.1^2)$	&$\mathcal{N}(1.0,0.3^2)$\\
 client bandwidth			& $bw_k$		&$\mathcal{N}(0.5,0.1^2)$	&$\mathcal{N}(1.0,0.3^2)$\\
 signal-noise ratio			& $SNR$			&1e2						&1e2		\\
 drop-out prob.				& $dr_k$		&$\mathcal{N}(\mathbb{E}[dr],0.05^2)$	&$\mathcal{N}(\mathbb{E}[dr],0.05^2)$\\
 region population			& $n_r$			&$\mathcal{N}(5.0,1.5^2)$	&$\mathcal{N}(50,15^2)$\\
 cloud-edge thrput.			& $BR$			&1e3						&1e3		\\
 max \# of rounds 			& $t_{max}$		&600						&400	   \\
 bits per sample 			& $BPS$			&6*8*8						&28*28*1*8 \\
 cycles per bit 			& $CPB$			&300						&400	   \\
 \# of local epochs 		& $\tau$		&5							&5		   \\
 loss function		 		& $f$			&MSE Loss					&NLL Loss	   \\
 learning rate 				& $\eta$		&1e-4						&1e-3		\\
 \hline
\end{tabular}
\label{tab:exp_setup}
\end{table}

In this paper, a cycle from the stage of model distribution, local training to model aggregation is called a \textit{federated round}. Note that the global aggregation is performed every federated round by FedAvg and our HybridFL, but HierFAVG performs it after several times of edge-level aggregation (i.e., it runs multiple federated rounds before a cloud aggregation). The cloud-level aggregation interval ($\kappa_2$ in reference \cite{hierFAVG}) for HierFAVG is set to 10, which is shown to be an optimal setting in their work.

For fair comparison, we ran all the protocols for the same number of (federated) rounds, denoted by $t_{max}$, which also means the same number of total epochs because each client runs the same number of local epochs, denoted by $\tau$, before the edge nodes conduct an edge-level aggregation.  

The length of a federated round, denoted by $T_{round}$, can be formulated as:

\begin{equation}
	T_{round} = T_{c2e2c} + \min \big\{T_{lim},\,\max_{k \in V_C'} \{ T_k^{comm} + T_k^{train} \} \big\}
\label{eq:T_round}
\end{equation}
where $T_{lim}$ is the preset limit of response time, which we configured as the time required by an extremely straggling client to finish its local training and communication with an average partition size. Given the performance (denoted by $s_k$) and bandwidth (denoted by $bw_k$) of the clients follow the normal distribution with the mean and standard deviation being $\mu$ and $\sigma$, respectively, the performance and bandwidth of such an extremely straggling client is set to be $\mu-3\sigma$.

Note that $V_C'$ represents the selected fraction of clients for FedAvg and HierFAVG, but for HybridFL we have $V_C' \equiv S(t)$ because of our quota-triggered aggregation. $T_{c2e2c}$ is the cloud-edge communication time, which is calculated by (\ref{eq:T_c2e2c}). $T_k^{comm}$ and $T_k^{train}$ are the communication time and local training time of client $k$, which are calculated by (\ref{eq:T_comm}) and (\ref{eq:T_train}), respectively.

\begin{equation}
	T_{c2e2c} = 3 \times \frac{msize \cdot m}{BR}
\label{eq:T_c2e2c}
\end{equation}
where $BR$ is the bit rate of the cloud-edge connection while for the edge-client wireless network we obtain its effective bit rate by applying the Shannon theorem to the corresponding bandwidth $bw_k$. The multiplicator "$3$" exists because the model upload typically spends twice as much time as the model download, given that uplink bandwidth is typically 50\% of the total. The size of the model ($msize$) is set to 5 MB and 10 MB for Tasks 1 and 2, respectively. For FedAvg, $T_{c2e2c} \equiv 0$ because it does not involve the edge layer.

\begin{equation}
	T_k^{comm} = 3 \times T_k^{download} = 3 \times \frac{msize}{bw_k \cdot \log(1+SNR)} 
\label{eq:T_comm}
\end{equation}

\begin{equation}
	T_k^{train} = \frac{|D_k^r| \cdot \tau \cdot BPS \cdot CPB}{s_k} 
\label{eq:T_train}
\end{equation}
The numerator in (\ref{eq:T_train}) quantifies the total number of CPU cycles needed for training the local partition $D_k^r$. 

Based on (\ref{eq:T_comm}) and (\ref{eq:T_train}) we can further model the energy consumed by end device for local training:

\begin{align}
	E_k & = E_k^{comm} + E_k^{train} \nonumber\\
		& = P_{trans} \cdot T_k^{comm} + P^{base}_{comp} s_k^3 \cdot T_k^{train} 
\label{eq:E}
\end{align}
where $P_{trans}$ is the power consumption of transmitter and $P^{base}_{comp} s_k^3$ represents the power for on-device computation based on the frequency power model \cite{my_power}. We set $P_{trans}$ and $P^{base}_{comp}$ to 0.5 and 0.7 Watt respectively based on the benchmarking results reported in ref. \cite{cell_power}.

\subsection{Experimental Results}
\begin{figure}[htb]
	\centering
	\includegraphics[width=250px]{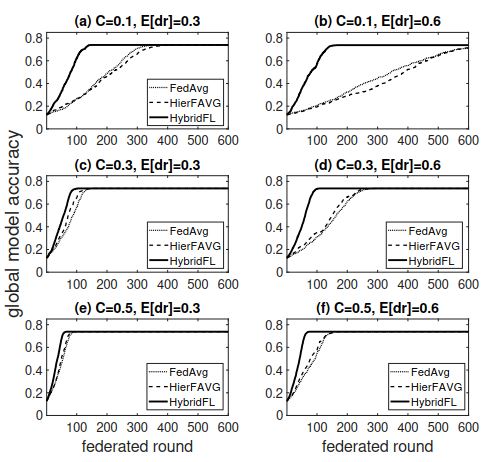}
	\caption{The trace of model accuracy during the FL process in Task 1: Aerofoil with $C=0.1, 0.3$ and 0.5. The cloud always keeps the best global model throughout the process. The drop-out probability of end devices follows the Gaussian distribution $\mathcal{N}(\mathbb{E}[dr],0.05^2)$.}
\label{fig:trace_task1}
\end{figure}

\begin{figure}[htb]
	\centering
	\includegraphics[width=260px]{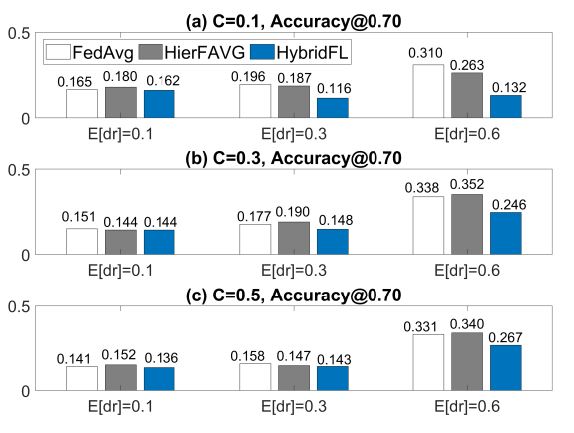}
	\caption{Comparing the energy consumption (Watt hours) of end devices when running Task 1 (Aerofoil) among FedAvg, HierFAVG and HybridFL. The values obtained are the average over all the end devices in the MEC system.}
\label{fig:energy_task1}
\end{figure}

\begin{figure}[htb]
	\centering
	\includegraphics[width=250px]{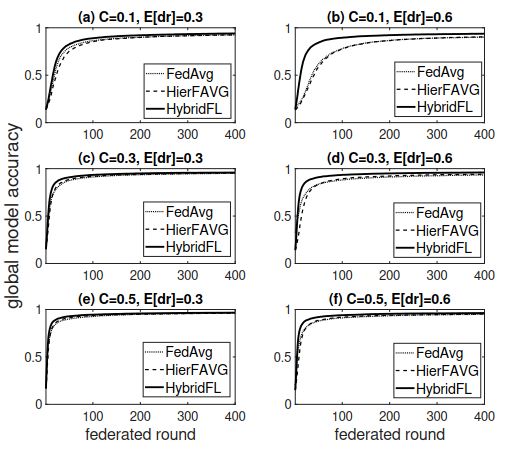}
	\caption{The trace of accuracy during the FL process in Task 2: MNIST with $C=0.1, 0.3$ and 0.5. The cloud always keeps the best global model throughout the process. The drop-out probability of end devices follows the Gaussian distribution $\mathcal{N}(\mathbb{E}[dr],0.05^2)$.}
\label{fig:trace_task2}
\end{figure}

\begin{figure}[htb]
	\centering
	\includegraphics[width=260px]{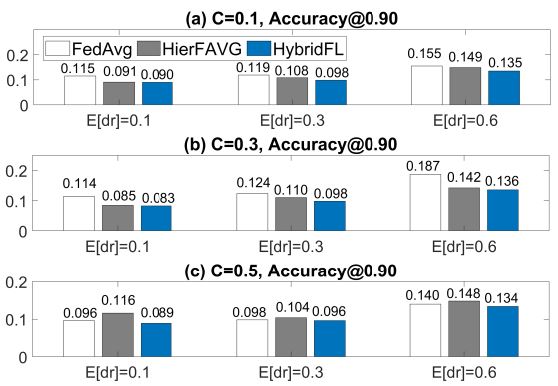}
	\caption{Comparing the average on-device energy (Watt hours) consumed for Task 2: MNIST in local training when using FedAvg, HierFAVG and HybridFL as the control protocols. The values obtained are the average over all the end devices in the MEC system.}
\label{fig:energy_task2}
\end{figure}

We ran the FL process in two ways: i) stop the process at a preset maximum round $t_{max}$, and ii) stop when a preset accuracy is achieved for the global model. In Table \ref{tab:res_task1} and Table \ref{tab:res_task2}, we present the results for task 1 and task 2, respectively, in terms of best model accuracy achieved, average round length (obtained when stopping at $t_{max}$), the number of rounds needed and the total time duration (for achieving the desired model accuracy). We also investigated the model convergence by comparing the accuracy traces (Figs. \ref{fig:trace_task1} and \ref{fig:trace_task2}) for FedAvg, HierFAVG and HybridFL. Figs. \ref{fig:energy_task1} and \ref{fig:energy_task2} show the average energy consumption by end devices.\\

\noindent \emph{Task 1: Aerofoil} 
Aerofoil is a numerical regression task. FL is performed to learn a global Fully-Connected Neural Network (FCN) model from a small group of clients that possess private aerofoil self-noise data\footnote{Airfoil Self-Noise Data Set, UCI. \url{https://archive.ics.uci.edu/ml/datasets/Airfoil+Self-Noise}}. Clients hold different partitions of the data without overlapping and cannot share the data with each other. This task simulates an industrial scenario where the production data are privacy-sensitive. The size of local partitions follow the Gaussian distribution specified in Table \ref{tab:exp_setup}.

% result table, task 1
\begin{table*}[ht]
\centering
\caption{Experimental results with Task 1: Aerofoil under different environmental settings of $\mathbb{E}[dr]$ and client selection proportion $C$.}
\begin{tabular}{c l c c c c c c c c c c c c}  % 14 cols
 \hline
 & & \multicolumn{6}{c}{Stop @$t_{max}$} & \multicolumn{6}{c}{Stop @Acc=0.70}\\
 & & \multicolumn{3}{c}{Best Accuracy} & \multicolumn{3}{c}{Round length (sec)} & \multicolumn{3}{c}{Rounds needed} & \multicolumn{3}{c}{Total time (sec)}\\
 $C$& & $0.1$	&$0.3$	&$0.5$ &$0.1$	&$0.3$	&$0.5$ &$0.1$	&$0.3$	&$0.5$ & $0.1$	&$0.3$	&$0.5$\\
 \hline
 \multirow{3}{*}{$\mathbb{E}[dr]=0.1$} & 
 	FedAvg 		& 0.727 & 0.727 & 0.727 & 52.42 & 73.12 & 80.73 & 238 & 95 & 56 & 12515.1 & 5585.2 & 4149.7\\
 &	HierFAVG 	& 0.727 & 0.726 & 0.728 & 51.56 & 71.90 & 81.08 & 250 & 80 & 50 & 14608.6 & 4765.8 & 4285.8\\
 &	HybridFL 	&\bf 0.729 & 0.727 & 0.728 &\bf 37.80 &\bf 63.80 &\bf 58.15 &\bf 113 &\bf 75 &\bf 49 &\bf 4254.6  &\bf 3341.4 &\bf 3143.0\\
 \hline
 \multirow{3}{*}{$\mathbb{E}[dr]=0.3$} & 
 	FedAvg 		& 0.728 & 0.727 & 0.727 & 64.21 & 83.64 & 87.90 & 376 & 125 & 74 & 25442.2 & 10674.6 & 6588.8\\
 &	HierFAVG 	& 0.727 & 0.728 & 0.727 & 66.74 & 83.24 & 88.14 & 340 & 130 & 60 & 22237.5 & 12025.9 & 6132.9\\
 &	HybridFL 	&\bf 0.729 & 0.728 &\bf 0.728 &\bf 38.94 &\bf 64.83 &\bf 69.84 &\bf 141 &\bf 77  &\bf 51 &\bf 7010.7  &\bf 4994.9  &\bf 3711.6\\
 \hline
 \multirow{3}{*}{$\mathbb{E}[dr]=0.6$} & 
 	FedAvg 		& 0.711 & 0.727 & 0.728 & 83.54 & 89.78 & 90.39 & 598 & 233 & 144 & 50122.4 & 21141.3 & 13108.4\\
 &	HierFAVG 	& 0.714 & 0.727 & 0.728 & 81.43 & 89.91 & 90.44 & 590 & 230 & 140 & 48922.2 & 21623.3 & 13565.6\\
 &	HybridFL 	&\bf 0.727 &\bf 0.728 & 0.728 &\bf 65.38 &\bf 73.23 &\bf 84.96 &\bf 160 &\bf 66  &\bf 62  &\bf 10584.1 &\bf 4780.4  &\bf 5488.3	\\
 \hline
\end{tabular}
\label{tab:res_task1}
\end{table*}

We ran 600 FL rounds to compare the best accuracy achieved and the average length of a round. The results are shown in the "Stop @$t_{max}$" column of Table \ref{tab:res_task1}. We can see that our protocol effectively shortens the average round length by 6\% to 42\% with slight improvements on the global model's accuracy in most cases. In Fig. \ref{fig:trace_task1}, we plot the trace of model accuracy over the FL process under the settings of $C \in \{0.1,0.3,0.5\}$ and $\mathbb{E}[dr] \in \{0.3,0.6\}$. From the figure we can observe a solid improvement in model convergence by HybridFL, especially under unstable MEC circumstances where end devices drop out frequently. In the setting of $\mathbb{E}[dr]$=0.6 and the selection proportion $C$=0.1 (Fig. \ref{fig:trace_task1}(b)), the global model can hardly converge in 600 rounds using FedAvg or HierFAVG, but reached its optimum in 200 rounds under the control of our HybridFL protocol.

% acc trace, task 1

We also tested the protocols by specifying a target model accuracy as the stop criterion, and observed the number of rounds needed for convergence and the total time duration. The results are shown in the "Stop @Acc" column of Table \ref{tab:res_task1}. HybridFL requires much fewer rounds and less time to achieve the accuracy target in Task 1, which yields up to 4$\times$ speed-up compared to FedAvg and HierFAVG. In the setting where the clients are mostly unreliable (i.e., $\mathbb{E}[dr]$=0.6), HybridFL can still achieve very fast convergence, requiring only about 1/3 of the rounds needed by HierFAVG. Another benefit of fast convergence is energy conservation. Fig. \ref{fig:energy_task1} shows the energy consumption of end devices. We can see that our protocol is most energy consumption friendly to end devices. HybridFL reduces the average energy usage of end devices by roughly 50\% for Task 1 in the case of $\mathbb{E}[dr]=0.6$ and $C=0.1$.

% energy, task 1

\bigskip
\noindent \emph{Task 2: MNIST}
In this task we aim to simulate a scenario in which the image samples are distributed over a relatively large fleet of end devices and they are not shared among the devices or allowed to be uploaded to the servers. This is a realistic scenario for mobile applications that are restricted by the privacy terms. In this experiment, 500 clients and 10 edge nodes are set up for running this task. Besides, to emulate the discrepancy in device users' behaviour (which leads to the biases in the data distribution over devices), we assigned the samples to clients by matching data labels with clients' indices -- sample $(x_i,y_i)$ has a 75\% chance to reside on (one of) the clients whose IDs are congruent to $y_i$ modulo 10 (the MNIST data set has 10 classes). This way the data distribution on each device is far from being IID. 

% result table, task 2
\begin{table*}[ht]
\centering
\caption{Experimental results with Task 2: MNIST under different environmental settings of $\mathbb{E}[dr]$ and client selection proportion $C$.}
\begin{tabular}{c l l l l l l l l l l l l l}  % 14 cols
 \hline
 & & \multicolumn{6}{c}{Stop @$t_{max}$} & \multicolumn{6}{c}{Stop @Acc=0.90}\\
 & & \multicolumn{3}{c}{Best Accuracy} & \multicolumn{3}{c}{Round length (sec)} & \multicolumn{3}{c}{Rounds needed} & \multicolumn{3}{c}{Total time (sec)}\\
 $C$& & $0.1$	&$0.3$	&$0.5$ &$0.1$	&$0.3$	&$0.5$ &$0.1$	&$0.3$	&$0.5$ & $0.1$	&$0.3$	&$0.5$\\
 \hline
 \multirow{3}{*}{$\mathbb{E}[dr]=0.1$} & 
 	FedAvg 	  & 0.936 & 0.958 & 0.962 & 377.28 & 378.02 & 378.02 & 200 & 66 & 32 & 75166.3 & 25327.3 & 12474.6\\
 &	HierFAVG & 0.936 & 0.958 & 0.964 & 377.65 & 378.26 & 378.26 & 150 & 40 & 30 & 60519.8 & 18911.3 & 15128.8\\
 &	HybridFL &\bf 0.940 &\bf 0.959 &\bf 0.965 &\bf 63.59  &\bf 96.55  &\bf 140.51 &\bf 124 &\bf 41 &\bf 26 &\bf 7856.9  &\bf 3762.9  &\bf 3730.6\\
 \hline
 \multirow{3}{*}{$\mathbb{E}[dr]=0.3$} & 
 	FedAvg 	  & 0.925 & 0.951 & 0.962 & 378.02 & 378.02 & 378.02 & 230 & 77 & 37 & 87322.3 & 29485.5 & 14364.7\\
 &	HierFAVG & 0.926 & 0.954 & 0.962 & 378.10 & 378.26 & 378.26 & 200 & 60 & 30 & 79432.7 & 26476.5 & 15128.8\\
 &	HybridFL &\bf 0.940 &\bf 0.959 &\bf 0.966 &\bf 109.72 &\bf 135.96 &\bf 113.20 &\bf 123 &\bf 41 &\bf 25 &\bf 15978.4 &\bf 7148.9  &\bf 4867.4\\
 \hline
 \multirow{3}{*}{$\mathbb{E}[dr]=0.6$} & 
 	FedAvg   & 0.901 & 0.933 & 0.950 & 378.02 & 378.02 &\bf 378.02 & 376 & 146 & 65 & 142513.1 & 55568.8 & 24949.2\\
 &	HierFAVG & 0.905 & 0.941 & 0.952 & 378.10 & 378.26 & 378.26 & 350 & 100 & 60 & 136171.6 & 41606.9 & 26476.5\\
 &	HybridFL &\bf 0.937 &\bf 0.960 &\bf 0.963 &\bf 37.59  &\bf 126.15 & 380.42 &\bf 118 &\bf 41  &\bf 31 &\bf 11743.1  &\bf 7334.8 &\bf 12171.8\\
 \hline
\end{tabular}
\label{tab:res_task2}
\end{table*}

We use the classic convolutional neural net LeNet-5 (consisting of two convolutional layers with max pooling and three fully connected layers) as the model for this image classification task. Again we ran FL for a fixed number of rounds first to observe the best accuracy and round length. The results are shown in the "Stop @$t_{max}$" column in Table \ref{tab:res_task2}, from which we can see that HybridFL outperformed FedAvg and HierFAVG in terms of the accuracy of the global model in all cases, especially when the participating devices are generally unreliable ($\mathbb{E}[dr]=0.6$). Fig. \ref{fig:trace_task2} tracks the accuracy of the global model in the training process under the settings of $C$= 0.1, 0.3 and 0.5 and $\mathbb{E}[dr]$= 0.3 and 0.6. It can be observed that the convergence of the global model is improved when using HybridFL as the controlling protocol. These results suggest that compared with FedAvg and HierFAVG, HybridFL can achieve the best global model in the fewest number of federated rounds.

% acc trace, task 2

We also compare the performance of HybridFL and the baseline protocols by specifying $acc$= 0.9 as the convergence target for the global model. The results are listed in the right part of Table \ref{tab:res_task2}. We can observe from the table that HybridFL significantly reduces the number of rounds and total time needed to achieve the accuracy target, compared with other two protocols. For example, HybridFL achieves a roughly 12$\times$ speed-up in the case where $\mathbb{E}[dr]$=0.6 and $C$=0.1, which represents a situation where the clients may drop out frequently and the participating fraction is restricted. 

Some interesting results were observed in the experiments. In both tasks 1 and 2 with $\mathbb{E}[dr]=0.6$, our protocol requires fewer rounds to converge given $C=0.5$ than that with $C=0.3$, but the total time consumption for $C=0.5$ is longer (see Tables \ref{tab:res_task1} and \ref{tab:res_task2}). This is because the extremely high drop-out probability (0.6 on average in the cases) of clients makes it almost impossible to engage 50\% of the them (given $C=0.5$) in training, even with the modulation of the regional slack factors. This is the case $|S(t)|<C\cdot n$. In such a case, the edge nodes and the cloud have to wait until the preset round-time limit is reached, and thus the round length is prolonged. To some extent, this observation explains why it is suggested in literature \cite{FL_survey1} that the selection proportion $C$ should not be set too large.

% energy, task 2

Device energy usage can be a key factor that affects the willingness of device owners to participate in the FL training. Fig. \ref{fig:energy_task2} shows the average energy consumption of an end device as a participant in the FL process to achieve the preset accuracy target 0.9 for Task 2. We find that the advantage of HybridFL in energy saving in Task 2 is not as prominent as in Task 1. Yet our protocol still managed to retain the on-device energy usage at the lowest level. This is because it enables much faster convergence (therefore less total training time for devices) than the baseline protocols. In practice, the energy-saving feature of HybridFL can help attract more end devices in each round.

%\subsection{Discussion}
The evaluation of the proposed protocol (HybridFL) with the two machine learning tasks under different environment settings demonstrates its effectiveness in terms of boosting the efficiency of FL, improving the global model's quality and saving on-device energy consumption in a three-layer MEC system. The reasons behind these improvements are three-fold. First, the quota-triggered regional aggregation in HybridFL effectively prevents the situation where some regions with extremely unreliable clients slow down the entire FL process. Second, we enable each edge node to modulate its regional quota based on its slack factor to improve the robustness of FL against client drop-out. Third, the cloud (i.e., global) aggregation is designed to be performed immediately after regional aggregation so that the global exchange of the model is made as early as possible.

\section{Conclusion}
\label{secV}
Thanks to the ever-increasing capacity of compute, storage and bandwidth at the edge of network, it has been a prominent trend that more and more end devices are infiltrated by the power of artificial intelligence. Meanwhile, the rising concerns about data privacy are changing the way we develop machine learning techniques and also reveal the great potential of using Federated Learning as a promising privacy-preserving solution. In this paper, we adapt FL to the mobile edge computing systems, aiming to improve both effectiveness and efficiency. We design a three-layer FL protocol called HybridFL to enable two levels of model aggregation to boost efficiency and mitigate the impact by the unreliable nature of end devices through modulating client selection in a region-wise manner, which results in a reasonable number of local updates as desired by the cloud. We conducted extensive experiments and the results demonstrate that HybridFL significantly improves FL in the MEC system by shortening the average length of a round, speeding up the convergence of the global model, promoting the model accuracy, and reducing device-side energy consumption.

In the future, we plan to extend our work to more complex system architectures which have different hierarchies and to diverse FL participants which have different roles. As another part of our future work, we plan to investigate how to improve the effectiveness of local training on each device without breaching the privacy constraints.

\section*{Acknowledgement}
This work is partially supported by the Worldwide Byte Security Information Technology Co. Ltd, Guangdong Project (Grant No. 2018B030325002), Key-Area Research and Development Program of Guangdong Province (Grant No.2020B010164003), Guangzhou Science and Technology Program Key Project (Grant Nos. 202007040002, 201902010040) and Guangzhou Development Zone Science and Technology (Grant No. 2018GH17).

% biography

% end bio

\newpage

% Appendices
% use appendices with more than one appendix
% then use \section to start each appendix
% you must declare a \section before using any
% \subsection or using \label (\appendices by itself
% starts a section numbered zero.)
%

\appendices
\section{Proof of Theorem 1*}
\label{apdx:prf}
First we define $\delta(t)$ as the weighted average of $\delta_k$ by $\gamma(k,r,t)$, which will be used later in the proof:

\begin{align}
	\delta(t) 	& \triangleq \sum_{k \in V_C} \gamma(k,r,t) \delta_k \nonumber \\
				& \leq \sum_{k \in V_C} \gamma(k,r,t) \bar{\delta} \nonumber \\
				& = \bar{\delta} \label{eq:delta}\\ 
				& \quad(\text{because} \sum_{k \in V_C} \gamma(k,r,t) = 1) \nonumber
\end{align}
where $\gamma(k,r,t)$ is the abbreviation of $\gamma(k,r(k),t)$ defined in (\ref{eq:reform-wt}) and $\delta_k$ is defined in (\ref{eq:delta_k}).

%Since we apply weights to client models in a different way from ref. \cite{IBM_FL}, we start the proof of our \textbf{Theorem} \ref{theorem1} based on the results provided in their paper (see Appendix B of \cite{IBM_FL}) by adapting it with $\delta(t)$ (defined in Eq. \ref{eq:delta}). First, the hypothetical model $w([z])$ is updated from $w([z-1])$ using gradient descent (on local devices) and our aggregation rule (Eq. \ref{eq:reform-wt}):
%
%\begin{equation}
%	w([z]) = w([z-1]) - \eta \sum_{k \in V_C} \gamma(k,r,t) \nabla F_k(w_k^{r(k)}([z-1]))
%\label{eq:w(z)_update}
%\end{equation}
%
%\begin{equation}
%	v_t([z]) = v_t([z-1]) - \eta \nabla F(v_t([z-1]))
%\label{eq:v(z)_update}
%\end{equation}

Combining (\ref{eq:w(z)_update}) and (\ref{eq:w_k(z)_update}), we have:
\begin{equation}
	w([z]) - w([z-1]) = -\eta\sum_{k \in V_C} \gamma(k,r,t) \nabla F_k(w_k^r([z-1]))
\label{eq:w_recur}
\end{equation}

From (\ref{eq:v(z)_update}) and (\ref{eq:w_recur}) given any $z \in ((t-1)\tau, t\tau]$, we have:
\begin{align}
	& \| w([z]) - v_t([z]) \| \nonumber \\
	& = \| w([z-1]) -\eta\sum_{k \in V_C} \gamma(k,r,t)\nabla F_k(w_k^r([z-1])) \nonumber \\ 
	&  \quad - v_t([z-1]) + \eta\nabla F(v_t([z-1]),t) \| \nonumber \\
	& = \| w([z-1]) -\eta \sum_{k \in V_C}\gamma(k,r,t) \nabla F_k(w_k^r([z-1])) \nonumber \\ 
	&  \quad - v_t([z-1]) + \eta \sum_{k \in V_C}\gamma(k,r,t) \nabla F_k(v_t([z-1])) \| \nonumber \\
	& = \| w([z-1]) - v_t([z-1])  \nonumber \\
	&  \quad - \eta \sum_{k \in V_C}\gamma(k,r,t) \big(\nabla F_k(w_k^r([z-1])) - \nabla F_k(v_t([z-1]))\big)  \|  \nonumber \\
	& \leq \| w([z-1]) - v_t([z-1]) \|  \nonumber \\
	&  \quad + \eta \sum_{k \in V_C}\gamma(k,r,t) \| \nabla F_k(w_k^r([z-1])) - \nabla F_k(v_t([z-1])) \| \nonumber \\
	& \qquad (\text{from triangle inequality}) \nonumber \\
	& \leq \| w([z-1]) - v_t([z-1]) \|  \nonumber \\
	& \quad + \eta \beta \sum_{k \in V_C}\gamma(k,r,t) \| w_k^r([z-1]) - v_t([z-1]) \| \nonumber \\
	& \qquad (\text{because } F_k(\cdot) \text{ is } \beta \text{-smooth})  \nonumber \\
	& \leq \| w([z-1]) - v_t([z-1]) \|  \nonumber \\
	& \quad + \eta \beta \sum_{k \in V_C}\gamma(k,r,t) \frac{\delta_k}{\beta} \big( (\eta\beta +1)^{z-1-(t-1)\tau} -1 \big) \nonumber \\
	& \qquad (\text{from [Lemma 2, ref. \cite{IBM_FL}})  \nonumber \\
	& = \| w([z-1]) - v_t([z-1]) \|  \nonumber \\
	& \quad + \eta \delta(t) \big( (\eta\beta +1)^{z-1-(t-1)\tau} -1 \big) \nonumber \\
	& \leq \| w([z-1]) - v_t([z-1]) \|  \nonumber \\
	& \quad + \eta \bar{\delta} \big( (\eta\beta +1)^{z-1-(t-1)\tau} -1 \big) \nonumber \\
	& \qquad (\text{from the definition of } \delta(t) \text{ in (\ref{eq:delta}}))
	\label{eq:thm1_prf0}
\end{align}

%Recalling the definitions of the hypothetical global model $w([z])$ and the auxiliary model $v_t([z])$ and considering the [(20), ref. \cite{IBM_FL}], for round $t$ (i.e., $z\in ((t-1)\tau, t\tau]$) we have:
%\begin{align}
%	& \| w([z]) - v_t([z])\| \nonumber \\
%	& = \Big\| w([z-1]) - \eta \sum_{k \in V_C} \gamma(k,r,t) \nabla F_k(w_k^r)([z-1])) \nonumber \\
%	& \qquad - v_t([z-1]) + \eta \nabla F(v_t([z-1])) \Big\| \nonumber \\
%	& \label{eq:thm1_prf0}
%\end{align}

Equivalently, we have:

\begin{align}
	&	\| w([z]) - v_t([z])\| - \| w([z-1]) - v_t([z-1])\| \nonumber \\
	&	\leq \eta \bar{\delta} \big( (\eta\beta +1)^{z-1-(t-1)\tau} -1 \big) \label{eq:thm1_prf1}
\end{align}

Since $w([z]) = v_t([z])$ when $z=(t-1)\tau$ according to our definition of the auxiliary model $v_t([z])$, we have $\| w([z])-v_t([z])\| = 0$ at $z=(t-1)\tau$. By summing up (\ref{eq:thm1_prf1}) over $z \in ((t-1)\tau, t\tau]$ (i.e., epochs in round $t$), we can derive:

\begin{align}
	&\| w([z]) - v_t([z])\| \nonumber \\
	&	= \sum_{i=(t-1)\tau +1}^z \| w([i]) - v_t([i]) \| - \| w([i-1]) - v_t([i-1])\| \nonumber \\
	&	\leq \eta \bar{\delta} \sum_{i=(t-1)\tau +1}^z \big( (\eta \beta +1)^{i-1-(t-1)\tau} -1 \big) \nonumber \\
	& 	= \eta \bar{\delta} \sum_{j=1}^{z-(t-1)\tau} \big( (\eta \beta +1)^{j-1} -1 \big)  \nonumber \\
	&	\quad \text{\;(let } j=i-(t-1)\tau \text{)} \nonumber \\
	&	= \eta \bar{\delta} \frac{(1-(\eta \beta+1)^{z-(t-1)\tau})}{-\eta \beta} - \eta \bar{\delta}(z-(t-1)\tau) \nonumber \\
	& 	= \frac{\bar{\delta}}{\beta}\big( (\eta \beta +1)^{z-(t-1)\tau} -1 \big) - \eta \bar{\delta}(z-(t-1)\tau) \nonumber \\
	&	= \bar{h}({z-(t-1)\tau}) \label{eq:thm1_prf2}
\end{align}

Recall that our target loss function $F(w,t)$ is $\rho$-Lipschitz (with regard to $w$) as a corollary from Assumption \ref{assump}. Using the result above we can further derive:

\begin{align}
	F(w([z]),t) - F(v_t([z]),t)	& \leq \| F(w([z])) - F(v_t([z])) \|  	\nonumber \\
								& \leq \rho \| w([z]) - v_t([z]) \|   	\nonumber \\
								& \leq \rho \bar{h}(z-(t-1)\tau)  \label{eq:thm1_prf3}
\end{align}

%\section{B}
%\label{apdx:b}

\end{document}